\documentclass[%
showpacs,preprintnumbers,
 amsmath,amssymb,
 aps,
 prd,
 lengthcheck,
letterpaper
]{revtex4-1}

\usepackage{graphicx}
\usepackage{hyperref}
\usepackage{ulem}
\usepackage{txfonts}
\usepackage{color}

\newcommand{\ket}[1]{\ensuremath{|{#1}\rangle}}

\begin{document}

\title{Stimulated neutrino transformation through turbulence on a changing density profile and application to supernovae}

\author{Kelly M.\ Patton}
\email{kmpatton@ncsu.edu}
\affiliation{Department of Physics, North Carolina State University,
Raleigh, North Carolina 27695-8202, USA}

\author{James P.\ Kneller}
\email{jim\_kneller@ncsu.edu}
\affiliation{Department of Physics, North Carolina State University,
Raleigh, North Carolina 27695-8202, USA}

\author{Gail C.\ McLaughlin}
\email{gail\_mclaughlin@ncsu.edu}
\affiliation{Department of Physics, North Carolina State University,
Raleigh, North Carolina 27695-8202, USA}

\begin{abstract}
{ We apply the model of stimulated neutrino transitions to neutrinos traveling through turbulence on a non-constant density profile. We describe a method to predict the location of large amplitude transitions and demonstrate the effectiveness of this method by comparing to numerical calculations using a model supernova (SN) profile. The important wavelength scales of turbulence, both those that stimulate neutrino transformations and those that suppress them, are presented and discussed. We then examine the effects of changing the parameters of the turbulent spectrum, specifically the root-mean-square amplitude and cutoff wavelength, and show how the stimulated transitions model offers an explanation for the increase in both the amplitude and number of transitions with large amplitude turbulence, as well as a suppression or absence of transitions for long cutoff wavelengths. The method can also be used to predict the location of transitions between antineutrino states which, in the normal hierarchy we are using, will not undergo Mikheev-Smirnov-Wolfenstein (MSW) transitions.  Finally, the stimulated neutrino transitions method is applied to a turbulent 2D supernova simulation and explains the minimal observed effect on neutrino oscillations in the simulation as as being due to excessive long wavelength modes suppressing transitions and the absence of modes that fulfill the parametric resonance condition.

}
\end{abstract}

\date{\today}

\maketitle

\section{Introduction}

Neutrino flavor transformations are a fascinating problem that have provided many interesting avenues of research over the years.  One such avenue is the effect different environments have on the evolution of neutrino flavor. 
In the case of a high neutrino density, such as in SN, the neutrinos will undergo collective oscillations due to interactions between the neutrinos themselves \cite{Qian:1995, Fuller:2006, Duan:2006, Balantekin:2007, Duan:2007, Duan:2006PRL, Duan:2010, Gava:2009, Esteban-Pretel, Banarjee:2011}. Also, neutrinos traveling through matter will undergo various MSW-type transitions due to interactions with the surrounding matter \cite{MS1986, Wolfenstein1977}. That matter profile may be very chaotic and there exists a rich history of study into the effect of fluctuations on the MSW-type transitions of neutrinos in various environments.  For example, the small fluctuations in the Sun and Earth have been investigated \cite{2001EPJC...20..507O,PhysRevD.43.2484, Nunokawa, Burgess, Sawyer}, as well as the large fluctuations in SN \cite{Loreti:1995ae, Friedland:2006ta, Fogli, Kneller:2010sc, borriello, Choubey,  Kneller:2012id}. Recently, collective effects in SN have been studied in the presence of density fluctuations \cite{Reid, Cherry, Cherry2} and the synthesis of collective, MSW effects and turbulence was investigated in \cite{2013PhRvD..88b3008L} where it was shown that features in the neutrino spectrum indicating the effects of self-interactions and/or MSW transitions 
may be partially or completely erased .  

Most of these works explained the effect of turbulence in terms of MSW-type transitions in the sense that density fluctuations increases the number of locations where a neutrino fulfills the MSW resonance condition in a particular mixing channel. In addition to increased mixing between the given two states, three flavor effects start to appear as the turbulence amplitude increases \cite{Kneller:2010sc, 2013arXiv1302.3825K} and oscillation channels that show no transitions with small amplitude turbulence are affected when the amplitude becomes large. In certain cases, a phenomenon known as depolarization has been observed, in which all information about the neutrinos' initial state is lost \cite{Loreti:1994ry, Kneller:2010ky}. Depolarization allows one to describe the final state of the ensemble of neutrinos as a simple probability distribution unrelated to the initial state.  It is not known for certain whether these conditions are met in SN. To determine if depolarization occurs in SN, it is necessary to study the histories of individual SN neutrinos. On the other hand, if depolarization is found not to occur in SN, the individual histories will be needed since the final state of the ensemble will depend on the initial conditions.  

All of this research has produced important advances, but in order to get the full picture we must have a complete understanding of the scales of fluctuations that affect the neutrino transformations most. And then there are some neutrino transitions in fluctuating matter densities which cannot be well explained by MSW theory because they occur of over a distance scale that is much longer than the oscillation scale. Instead, they fit into a category of transformations known as stimulated transitions, caused by parametric resonance.  This effect is similar to MSW-type transitions, in that it results from interactions with matter.  However, whereas MSW-type transitions occur when neutrinos pass through material at a resonant density, stimulated transitions are caused by matching the frequency of fluctuation to the eigenvalue splitting of the neutrino system \cite{Friedland:2006ta, Ermilova, 1987PhLB..185..417S, Akhmedov, 1989PhLB..226..341K, 2009PhLB..675..69K}. This frequency matching can happen at any density scale, allowing for the possibility of flavor transformations even far from the MSW resonance.  It has also been shown that harmonics of the eigenvalue splitting can cause a parametric resonance \cite{Kneller:2012id}. Parametric resonances in neutrinos are very similar to stimulating transitions between energy levels in an atom using a laser \cite{Kondo:1992}, and in fact many of the equations used to describe the two systems closely resemble one another.  Since any density profile, even a turbulent one, can be decomposed into Fourier modes of different frequencies, the parametric resonance approach is a natural one to take.  

In previous work \cite{Patton:2013}, we have shown that the approach based upon the rotating wave approximation (RWA) can be used to accurately predict the transition wavelength and amplitude for neutrinos traveling through turbulence with up to forty modes.  This previous work was restricted to turbulence superimposed on a constant density profile and the purpose of this present work is to expand to tackling the issues associated with a changing profile.  
In Section \ref{Methods}, we reiterate the RWA solution and the important scales of the problem and propose how one can apply these results to a non-constant density profile. In Section \ref{Predict} we then test our proposal by applying it to a model SN profile and discover that it works well. Based upon that success, section \ref{Vary} examines the effect of changing parameters of the power spectrum of the turbulence and then in section \ref{AntiNu} we show how we are able to predict transition between antineutrino states, clearly illustrating the utility of the stimulated transition approach to explain those cases where MSW transitions are not expected. In section \ref{Application} we use our method to explain the recent null result of turbulence upon neutrinos passing through a 2D simulation, and then finish with our conclusions. 

\section{Analytic Solution}\label{Methods}

\subsection{Constant Density}

We are interested in calculating the probability that a neutrino starting in state $\ket{\nu(r)}$ at point $r$ is in state $\ket{\nu'(r')}$ at some later point $r'$.  These states are both at different locations and in different Hilbert spaces. This can be accomplished by computing the $S$-matrix, which provides a connection between the initial and final states through the equation $\ket{\nu'(r')} = S \, \ket{\nu(r)}$ \cite{Kneller:2006,Kneller:2009}.  This $S$-matrix is the path ordered exponential of the Hamiltonian and evolves via the equation
\begin{equation}\label{eq:dSdr}
\imath \frac{dS}{dr} = H \, S.
\end{equation}
In the flavor basis, the Hamiltonian $H$ is given by $H^{(f)} = U_{0} K^{(m)} U_{0}^{\dag} + V^{(f)}$, where $U_{0}$ is the vacuum mixing matrix, $K^{(m)}$ is a matrix containing the vacuum eigenvalues, and $V^{(f)}$ describes some potential through which the neutrino moves.  We separate our potential into two pieces: a smooth base potential $\breve{V}$ and the perturbation $\delta V$.  With this separation, we now write our Hamiltonian as $H=\breve{H} + \delta V$.  The Hamiltonian $\breve{H}$ and all other quantities denoted by a `breve' are unperturbed quantities, which includes the vacuum contribution and the smooth potential, but not the perturbation $\delta V$.  We can define an unperturbed matter basis, which is related to the flavor basis through a unitary mixing matrix $\breve{U}$ by $\breve{H}^{(f)} = \breve{U} \breve{K}^{(\breve{m})} \breve{U}^{\dag}$, where $\breve{K}^{(\breve{m})}$ is a diagonal matrix of the eigenvalues of $\breve{H}$.

At this point, we will make the assumption that the potential $V$ has only one non-zero component, that of $V_{ee}$, making the potential `MSW'-like. We are interested in the effects of perturbations of the form
\begin{equation}\label{eq:perturbation}
\delta V(r) = \breve{V}(r) \sum_{a=1}^{N_{k}}\left[ C_{a} \sin{(q_{a} r + \eta_{a})}\right],
\end{equation}
where $N_{k}$ is the number of modes, with each mode having wavenumber $q_{a}$, amplitude $C_{a}$, and phase shift $\eta_{a}$.  The phase shifts $\eta_{a}$ are randomly distributed between 0 and $2\pi$, while the amplitudes and wavenumbers follow a Kolmogorov power spectrum, given by\cite{2013arXiv1302.3825K}
\begin{equation}
E(q) = \frac{(\alpha - 1)}{2 q_{cut}}\left(\frac{q_{cut}}{|q|}\right)^{\alpha} \Theta(|q| - q_{cut}), 
\end{equation}
where $\alpha = 5/3$ and $q_{cut}$ is the cutoff scale and sets the lowest allowed frequencies for the spectrum.  The power spectrum $E(q)$ is a measure of the power in the turbulence at each frequency and is related the amplitudes $C_{a}$ in equation (\ref{eq:perturbation}) through the simple relationship $E(q_{a}) \propto C_{a}^{2}/q_{a}$.  Due to the inverse power law nature of the spectrum, longer wavelength modes will have more power and thus larger amplitudes.  The amplitudes and frequencies for the $N_{k}$ modes are chosen using the Randomization Method described in \cite{2013PhRvD..88b3008L, 2013PhRvD..88d5020K,2013arXiv1302.3825K}.  
 
 In this work, we investigate the effects of changing two parameters of the turbulent spectrum.  The first is the root-mean-square (RMS) amplitude, defined as $C_{rms}^{2} = \sum_{a}C_{a}^{2}$.  SN simulations have reported a wide range of RMS amplitudes.  The amount of turbulence in the SN progenitor has been reported as $C_{rms}\sim 10^{-5}$, providing a lower limit to the turbulence in the explosion \cite{Meakin:2007}.  Upper limits in SN turbulence reach up to $C_{rms}\sim 0.3-0.5$, justified from the multidimensional physics required for a large star to actually explode in simulations \cite{2013PhRvD..88b3008L}.  Given the variety of values for $C_{rms}$ reported, we study how these changes alter the neutrino propagation.

The second parameter is the cutoff wavelength, defined as $\lambda_{cut} = 2\pi \hbar c/q_{cut}$.  The cutoff wavelengths used in previous work have varied from tens of kilometers \cite{Reid} to thousands or tens of thousands of kilometers \cite{Burgess, Sawyer, 2013PhRvD..88b3008L, Kneller:2010sc, 2013arXiv1302.3825K, 2013PhRvD..88d5020K}.  From the work in \cite{Patton:2013}, we know that long wavelength modes can have a suppressive effect on the neutrino transformations, so a change in cutoff wavelength could make a drastic difference in the resulting transitions.  Once again, due to the large range of previously used cutoff wavelengths, we look into how varying this parameter affects the neutrino propagation.  

In the case of a constant density ($\breve{V}(r) = V_{0}$), the probability of transition between the unperturbed states 1 and 2, the eigenstates of $\breve{H}$, is given by \cite{Patton:2013}
\begin{equation}
P_{12} = \frac{\kappa^{2}}{Q^{2}} \sin^{2}{(Q r)},
\end{equation}
where 
\begin{eqnarray}\label{eq:kappa}
\kappa  & = &  \frac{\breve{U}_{e1}^{*} \breve{U}_{e2}}{|\breve{U}_{e1}|^2 - |\breve{U}_{e2}|^2}
\left(\sum_{a=1}^{N_k} n_{\star a}\,q_{a} \right) \nonumber \\
& &  \times  \prod_{a=1,}^{N_k} (-\imath)^{n_{\star a}}\,J_{n_{\star a}}(z_{a})
\,\exp\left[\imath\left(n_{\star a}\,\eta_a +z_{a} \cos\eta_a \right)\right],
\end{eqnarray}
and $Q^{2} = p^{2} + \kappa^{2}$.  The oscillation wavelength $Q^{-1}$ is an important quantity for determining where transitions occur on a changing profile, as will be explained later.  We further define $2 p = \delta \breve{k}_{12} + \sum_{a} n_{\star a} q_{a}$ and $z_{a} = C_{a} V_{0} (|\breve{U}_{e1}|^2 - |\breve{U}_{e2}|^2)/q_{a}$.  Here, $\delta \breve{k}_{12} = \breve{k}_{1} - \breve{k}_{2}$, where the $\breve{k}_{i}$ are the eigenvalues of $\breve{H}$. The full derivation of this RWA solution can be found in \cite{Patton:2013}.    

The $n_{\star a}$ in the definition of $p$ are integers, chosen using a simplex with $-\kappa^{2}/Q^{2}$, or negative the amplitude, as the potential. The simplex locates the set of $n_{\star a}$ which best satisfy the parametric resonance condition $|\delta \breve{k}_{12} + \sum_{a}n_{\star a}q_{a}| \approx 0$ while also minimizing the value of $-\kappa^{2}/Q^{2}$.  This method was shown to produce predictions similar to brute force numerical results in \cite{Patton:2013}.   

Using this analytic solution, two important wavelength scales have been identified.  The first is associated with the neutrino mass splitting and defined as 
\begin{widetext}
\begin{equation}\label{eq:lambdaSplit}
 \lambda_{fluct, split}
  \sim 20\, {\rm km} \, \left[ \left( \left(\frac{\delta m^{2}}{3\times10^{-3}\,{\rm eV^{2}}}\right)\,\left(\frac{20 {\rm MeV}}{E}\right)\, \left(\frac{\cos 2\theta}{0.95}\right) - 0.53 \left( \frac{\rho}{1000 \, {\rm g/cm^{3}}}\right) \right)^{2} + \, 0.1\left(\left(\frac{\delta m^{2}}{3\times10^{-3}\,{\rm eV^{2}}}\right)\,\left(\frac{20 {\rm MeV}}{E}\right)\, \left(\frac{\sin 2\theta}{0.3}\right)\right)^{2} \right]^{-1/2}.
\end{equation}
\end{widetext}
 Density perturbations with wavelengths close to $\lambda_{fluct,split}$ cause parametric resonances and have been discussed in works such as \cite{Friedland:2006ta, Ermilova, 1987PhLB..185..417S, Akhmedov, 1989PhLB..226..341K, 2009PhLB..675..69K, Kneller:2012id}.  This length scale corresponds to those modes with a non-zero contribution to the parametric resonance condition, $|\delta \breve{k}_{12} + \sum_{a}n_{\star a}q_{a}| \approx 0$, where then $\lambda_{fluct,split} \approx 2\pi \hbar c/q_{a} \approx 2\pi\hbar c/\delta \breve{k}_{12}$.
 
 The second wavelength is related to the amplitude of the density fluctuations, and can suppress the transitions caused by $\lambda_{fluct,split}$.  A good physical analog this suppression is the Stark effect.  For a two-level atom, a laser tuned to the correct frequency can stimulate a transition between states.  Applying an external field to that atom will shift the wavefunctions and energies of the original states, such that the laser that once stimulated transitions can no longer do so.  

In the neutrino system, the ``laser'' consists of the modes which contribute to the parametric resonance.  The ``external electric field'' is made up of the rest of the modes in the turbulence.  These modes cause a distortion to the system which prevents the parametric resonance from causing a transition.  We showed in \cite{Patton:2013} that this distortion can be quantified, and that the modes which cause the largest distortion have wavelengths given by $\lambda_{fluct, ampl}$, given by 
 \begin{widetext}
 \begin{equation}\label{eq:FluctAmpl}
 \lambda_{fluct, ampl} \sim 125\, {\rm km} \, \left( \frac{0.1}{C_{a}}\right) \, \left( \frac{1000 \, {\rm g/cm^{3}}}{\rho}\right) \frac{\sqrt{\left( \left(\frac{\delta m^{2}}{3\times10^{-3}\,{\rm eV^{2}}}\right)\,\left(\frac{20 {\rm MeV}}{E}\right)\, \left(\frac{\cos 2\theta}{0.95}\right) - 0.53 \left( \frac{\rho}{1000 \, {\rm g/cm^{3}}}\right) \right)^{2} + \, 0.1\left(\left(\frac{\delta m^{2}}{3\times10^{-3}\,{\rm eV^{2}}}\right)\,\left(\frac{20 {\rm MeV}}{E}\right)\, \left(\frac{\sin 2\theta}{0.3}\right)\right)^{2}}}{\left( \left(\frac{\delta m^{2}}{3\times10^{-3}\,{\rm eV^{2}}}\right)\,\left(\frac{20 {\rm MeV}}{E}\right)\, \left(\frac{\cos 2\theta}{0.95}\right) - 0.53 \left( \frac{\rho}{1000 \, {\rm g/cm^{3}}}\right) \right)}.
 \end{equation}
 \end{widetext}
Note that equation (\ref{eq:FluctAmpl}) has been modified: here we have included mixing matrix elements that were approximated as being of order 1 previously.  The values of $\lambda_{fluct,ampl}$ correspond to those modes where the value of $z_{a} = C_{a} V_{0} (|\breve{U}_{e1}|^2 - |\breve{U}_{e2}|^2)/q_{a}$ is close to a zero of the Bessel function $J_{0}(z_{a})$.  When such a zero is encountered, the value of $\kappa$, and the amplitude of the transition, is heavily suppressed.  The modes that contribute the $J_{0}(z_{a})$ to the induced wavelength and amplitude have $n_{\star a}=0$ and therefore are not the modes that trigger the parametric resonance condition.  However, they distort the underlying density profile.

\begin{figure}[t!]
\includegraphics[width=.9\linewidth]{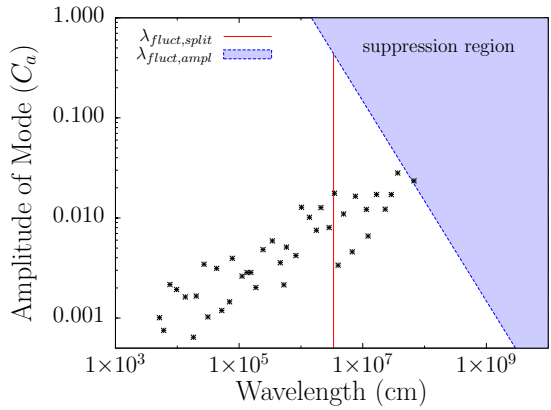}
\caption{Example of the important scales and suppression region in terms of wavelength and amplitude of modes in the density perturbation imposed on a constant density of $\rho = 941$ g/cm$^{3}$.  The solid red line indicates $\lambda_{fluct,split}$, while the dashed blue curve shows $\lambda_{fluct,ampl}$.  The shaded blue area is the suppression region.  Black points indicate amplitudes and wavelengths for a Kolmogorov power spectrum known to produce suppression.  One of the modes in this spectrum does fall in the suppression region.}\label{fig:lambdaVsC}
\end{figure}

\begin{figure}[t!]
\includegraphics[width=.9\linewidth]{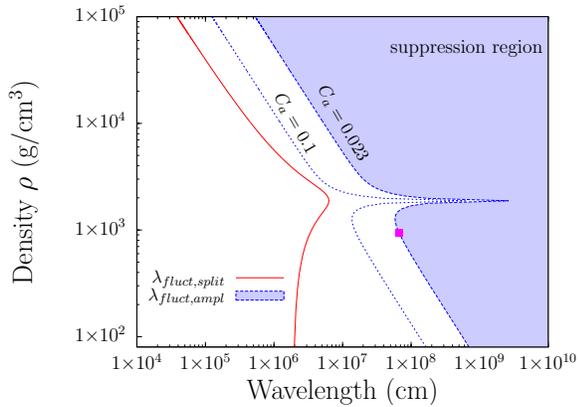}
\caption{Example of the important scales and suppression region in terms of wavelength and density.  The red curve, as in figure \ref{fig:lambdaVsC}, shows the value of $\lambda_{fluct,split}$ while the dashed blue curves show the value of $\lambda_{fluct,ampl}$ for $C_{a} = 0.1$ and $C_{a}=0.023$. The magenta point is the longest wavelength mode from the example spectrum used in figure (\ref{fig:lambdaVsC}), which has amplitude $C_{a} = 0.023$. }\label{fig:lambdaVsRho}
\end{figure}

These results are summarized in figures (\ref{fig:lambdaVsC}) and (\ref{fig:lambdaVsRho}).  Figure (\ref{fig:lambdaVsC}) shows the amplitude as a function of perturbation wavelengths at a constant density of 941 g/cm$^{3}$, which is half of the MSW density for neutrinos with an energy of 20 MeV, $\delta m^{2}$ = 3$\times10^{-3}$ eV$^{2}$ and a mixing angle of $9^{\circ}$.  The solid red line shows the value of $\lambda_{fluct,split}$, while the dashed blue line represents the second scale of $\lambda_{fluct,ampl}$.  The shaded region above $\lambda_{fluct,ampl}$ is the suppression region.  Fluctuations whose wavelength and amplitude places them in the suppression region are likely to cause a suppression of the stimulated transitions.  The black points shown on this figure are the amplitude and wavelengths for a randomly generated Kolmogorov spectrum with RMS amplitude of 10\% and a cutoff wavelength of $314.1$ km.  This particular set of wavelengths and frequencies was shown to exhibit a dramatic suppression effect in figure (4) of \cite{Patton:2013}, reducing the amplitude of the transitions from unity to $\sim 0.1$.  In fact, as we can see in figure (\ref{fig:lambdaVsC}), some of the points in this spectrum fall very close to or within the suppression region. 

We can also look at these scales in the plane of perturbation wavelength and density, shown in figure (\ref{fig:lambdaVsRho}). As before, the solid red and dashed blue curve indicate $\lambda_{fluct,split}$ and $\lambda_{fluct,ampl}$, respectively, and the shaded blue area is the suppression region. In this case, multiple lines can be drawn for $\lambda_{fluct,ampl}$ depending on the amplitude $C_{a}$ being considered.  At a given density, any perturbation with an amplitude $C_{a}$ and a wavelength that places it above the $\lambda_{fluct,ampl}$ curve for that $C_{a}$ will cause suppression. As an example, we have plotted the curve for $C_{a} = 0.023$, the amplitude of the first mode in the spectrum plotted in figure (\ref{fig:lambdaVsC}).  The single magenta point shows the wavelength and density for that mode in this example.  We already know this particular spectrum displays suppression, and now we can confirm that this mode specifically is contributing to that suppression as it falls on the $\lambda_{fluct,ampl}$ curve.

\subsection{Stimulated Transitions with Non-Constant Density Profiles}

We want to now move to the case of a non-constant $\breve{V}$.  As a first case we consider a smooth parameterized supernova model taken from Fogli \textit{et al.}, with the shock at the $t=4$s position \cite{2003PhRvD..68c3005F}.  We use a two-flavor system for simplicity, with the energy of the neutrino set to 20 MeV, a vacuum mass splitting of $\delta m^{2} = 3\times10^{-3}$ eV$^{2}$ and a vacuum mixing angle of $\theta=9^{\circ}$.  This parametrized SN model includes a single shock and a 20 MeV neutrino will pass through three MSW resonances, one of which is the shock, for the mixing parameters and neutrino energy chosen.   The turbulence in front of the shock at 4.22$\times10^{4}$ km is expected to be negligible, so our calculation ends at the shock in all cases discussed here.  

\begin{figure}
\includegraphics[width=.9\linewidth]{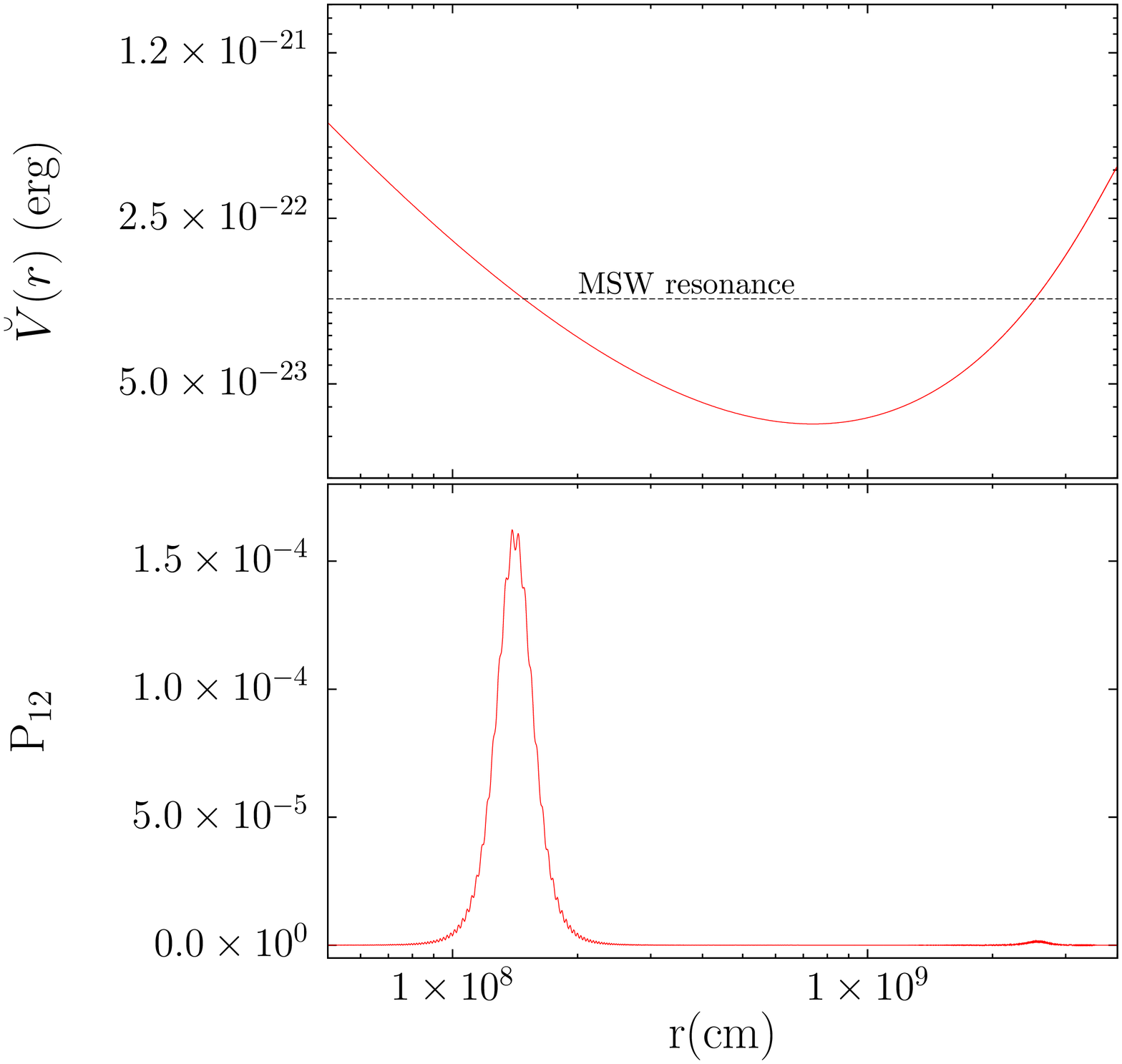}
\caption{The matter potential $\breve{V}$ (top) and probability of transition (bottom) for the Fogli \textit{et al.} profile with no perturbations.  The MSW resonance potential for neutrinos with an energy of 20 MeV, $\delta m^{2} = 3\times10^{-3}$ eV$^{2}$, and $\theta = 9^{\circ}$ is indicated in the top panel by the dotted black line.  The very low probability indicates that any large transitions seen with perturbations are due to parametric resonances.}\label{fig:fogliNoPert}
\end{figure}

The probability $P_{12}$ in the unperturbed matter basis for the unperturbed potential is shown in figure (\ref{fig:fogliNoPert}).  The small rises in the probability near $1\times10^{8}$ cm and $2\times10^{9}$ cm  occur at the first two MSW resonances.  Based on these results, we can conclude that any large transitions in probability that we see after inserting turbulence must be due to parametric resonances caused by the perturbations.

Our proposal for how to apply the constant density method is to insert spatial dependence into previously constant quantities, such as $\breve{U}$ and $\delta \breve{k}$, and calculating the stimulated transition wavelength $1/Q$ as a function of position.  We then determine whether the transition wavelength is less than the density scale height, defined as $r_{\rho} = |\, \rho(r)/(d\rho/dr)\,|$, where the mass density $\rho(r)$ is proportional to $\breve{V}_{ee}(r)$.  The density scale height is a measure of how fast the potential changes with distance. In order to see a transition we require that the predicted transition wavelength, $1/Q$, must be smaller than $r_{\rho}$.  This indicates that the transition is occurring faster than the density is changing. In the following sections, we will use this method to make predictions of the locations of transitions.  We will also use this method to study the effects of changing the turbulence parameters $C_{rms}$ and $\lambda_{cut}$, as well as predict transitions between antineutrino states.

\section{Predicting Transitions}\label{Predict}

\subsection{Four Modes}

One of the major advantages of the RWA approach is the ability to predict where large transitions are likely to occur as neutrinos travel through a density profile.  Numerical calculations can be computationally expensive, but with the RWA approach, we can determine if a profile will prove interesting or not before using that valuable computing power.  Here, we will show the predictive power of the RWA solution.  

We will start with a case of four modes ($N_{k}=4$) with a RMS amplitude of $C_{rms}=0.1$ and a cutoff wavelength of $\lambda_{cut}=314.1$ km.  Frequencies, amplitudes and phases for the four modes are listed in Table (\ref{table:4modes}).  The small number of modes was chosen to make transitions very clear to the reader while the cutoff wavelength ensures that no suppressive effects will be present.  The method also works with higher numbers of modes, and an example of this will be described later.  There are two steps involved in predicting transitions: finding parametric resonances and comparing predicted wavelengths to the density scale height.

\begin{table}[t]
\caption{Values of $q_{a}$, $C_{a}$ and $\eta_{a}$ used for the $N_{k}$ = 4 modes example shown in figure \ref{fig:4modes} with $\lambda_{cut} = 314.1$ km and $C_{rms} = 0.05$. Figure \ref{fig:CrmsVary} was created by multiplying each $C_{a}$ by a constant factor (eg. for $C_{rms}$ = 0.1, each $C_{a}$ was multiplied by 2).}
\begin{center}
\begin{tabular}{cccc}
 \hline
  \hline
 a & q$_{a}$ (erg) & C$_{a}$ & $\eta_{a}$\\
 \hline
 1 &  $1.48355\times10^{-23}$ &  0.07395 & 4.4432\\
 2 & $8.16779\times10^{-23}$ &  0.02167 & 1.7732\\
 3 & $1.77399\times10^{-21}$ &  0.02522 & 0.07516 \\
 4 & $5.38203\times10^{-20}$ &  0.006297 & 5.9402\\
 \hline
 \hline
\end{tabular}
\end{center}
\label{table:4modes}
\end{table}%

\begin{figure}
\begin{centering}
\includegraphics[width=.8\linewidth]{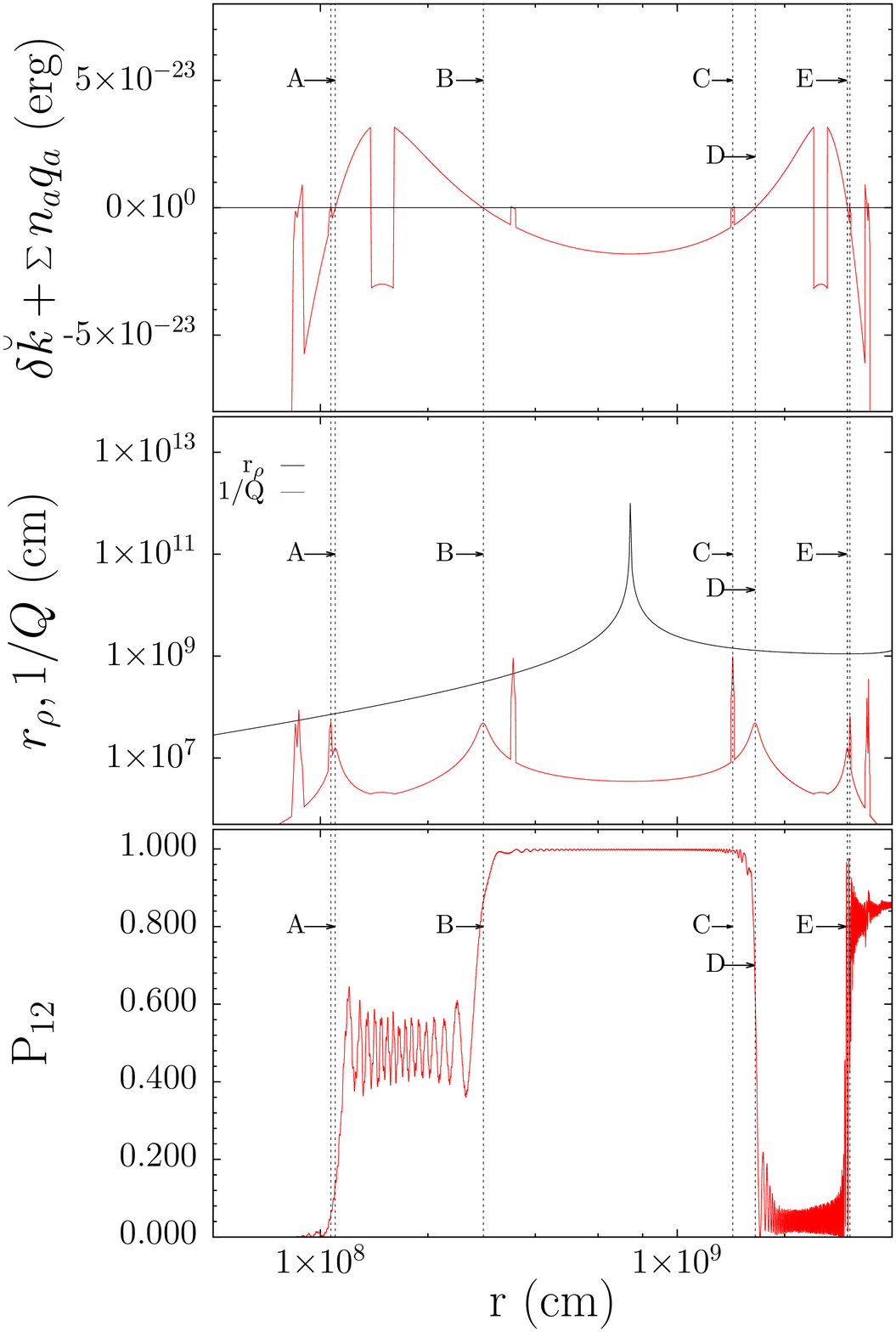}
\caption[Example of the transitions caused by a perturbation with four modes.]{Example of the transitions caused by a perturbation with four modes. (Top) The sum $\delta\breve{k} + \sum n_{a}q_{a}$ (red) is used to locate parametric resonances.  This condition is met when the sum equals 0 (black).  (Middle)  The predicted wavelengths from the RWA solution (red) compared to the density scale height (black).  (Bottom)  Probability of transition $P_{12}$ in the unperturbed matter basis.  Predicted transitions, located by the method described in the text, are indicated by vertical dashed lines $A$ through $E$ in all panels.}\label{fig:4modes}
\end{centering}
\end{figure}

We first determine the RWA solution at each point along the profile using the simplex method. Then, we find all locations where the parametric resonance condition is met ($0\approx \delta\breve{k} + \sum n_{a}q_{a}$).  Large transitions are only possible when this condition is satisfied.  An example of this is shown in the top panel of figure (\ref{fig:4modes}).  The red curve shows the sum $\delta\breve{k} + \sum n_{a}q_{a}$, and the solid black line is at 0.  There are many locations where this condition holds, but a large fraction will be eliminated by the next step.  

Next, we must determine if the constant density approximation is valid.  As mentioned above, we will use the density scale height $r_{\rho}$ to accomplish this.  The RWA solution found through the simplex method provides us with predicted wavelengths at each point along the density profile.  Since we have already located points where parametric resonances occur, we focus on the predicted wavelengths at those locations.  If the predicted wavelength is lower than $r_{\rho}$ the constant density approximation holds and we can expect a transition.  This is illustrated in the middle panel of figure (\ref{fig:4modes}).  The red curve shows the predicted wavelengths while the solid black curve is the density scale height for our chosen base profile.  Requiring that the predicted wavelength be less than the density scale height eliminates many points that fulfill the parametric resonance.  After these two steps, we are left with five potential transition locations, indicated by vertical dashed lines labeled $A$ through $E$. 

Finally, in the bottom panel of figure (\ref{fig:4modes}), we see the numerically calculated probability for this case.  At each point $A$ through $E$, a substantial transition is observed.  The method outlined above accurately predicted the locations of each of these transitions.  All of the large transitions in this example are explained by parametric resonance and the RWA solution.

In contrast to the unperturbed case in figure (\ref{fig:fogliNoPert}) which hit a maximum probability of less than $2\times10^{-4}$, the transitions here are between 50\% and 100\%.  Transitions such as $A$ and $B$, which do not reach 100\%, are due to the predicted wavelength being longer than the distance over which the transition occurs.  For instance, transition $A$ has a predicted wavelength of $\sim 1.6\times10^{7}$ cm.  Before the neutrino can travel that distance, however, the parametric resonance condition is no longer met and the transition ends.  An extreme example of this is transition $C$.  The parametric resonance condition at location $C$ is met for an almost negligible distance.  The system barely begins transitioning before moving off the resonance.  Thus, even though location $C$ meets our criteria for a potential transition, nothing observable occurs there.  

\subsection{20 Modes}

\begin{figure}
\begin{centering}
\includegraphics[width=.8\linewidth]{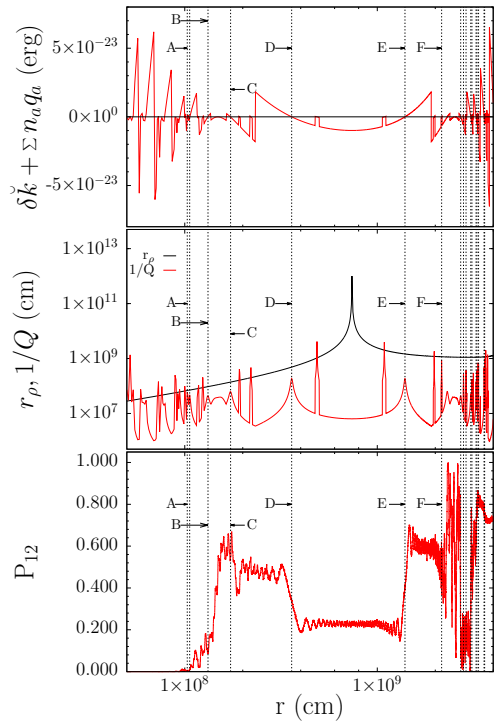}
\caption[Example of transitions caused by a perturbation with 20 modes.]{Same as figure (\ref{fig:4modes}), but for a perturbation with 20 modes.}\label{fig:20modes}
\end{centering}
\end{figure}

\begin{figure}
\begin{centering}
\includegraphics[width=.8\linewidth]{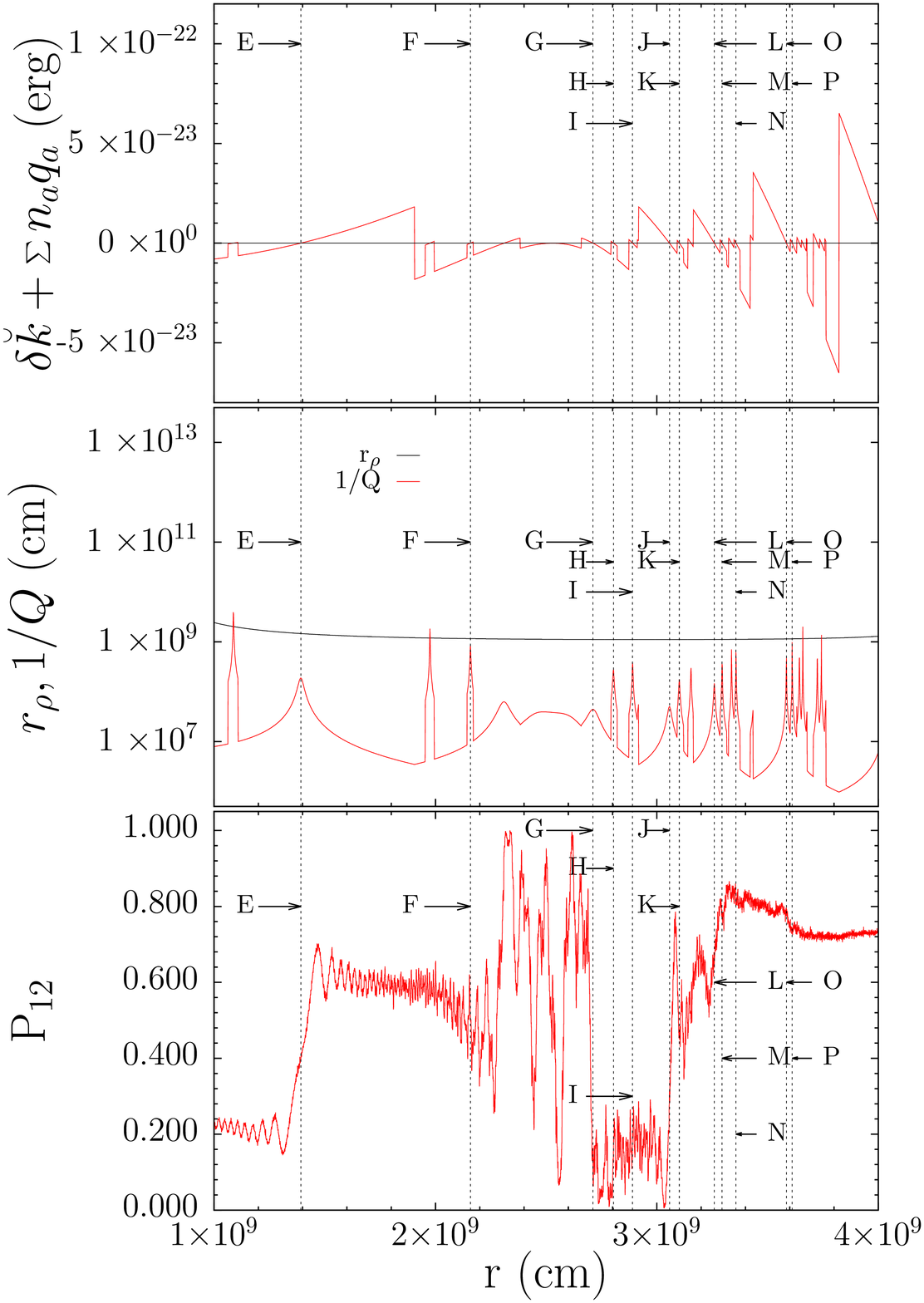}
\caption[Close up of the last $3\times 10^{4}$ km of propagation for the example in figure (\ref{fig:20modes})]{A close up of the last $3\times 10^{4}$ km of propagation for the example shown in figure (\ref{fig:20modes}).}\label{fig:20modesClose}
\end{centering}
\end{figure}

A realistic turbulent profile would have many more than four modes.  Figures (\ref{fig:20modes}) and (\ref{fig:20modesClose}) show an example with 20 modes, showing that this method for predicting transitions still works when the number of modes is increased.  The turbulent spectrum has an RMS amplitude $C_{rms} = 0.05$, half that of the case used in figure (\ref{fig:4modes}), and a cutoff scale of $\lambda_{cut} =314.1$ km.  Values for the frequencies, amplitudes and phases are included in table (\ref{table:20modes}). As in the previous figure, the top panel shows the sum $\delta\breve{k} + \sum n_{a}q_{a}$ (in red). The middle panel shows the predicted wavelengths in red as compared to the density scale height in black, and the bottom panel shows the probability $P_{12}$.  Locations which meet both conditions described above are indicated with vertical dashed lines.  

\begin{table}[t]
\caption{Values of $q_{a}$, $C_{a}$ and $\eta_{a}$ used for the $N_{k}$ = 20 modes example shown in figures \ref{fig:20modes} and \ref{fig:20modesClose} with $\lambda_{cut} = 314.1$ km and $C_{rms} = 0.05$. Figure \ref{fig:kCutVary} was created by multiplying each $q_{a}$ by a constant factor (eg. for $\lambda_{cut} = 3141 km$, each $q_{a}$ was multiplied by 0.1).}
\begin{center}
\begin{tabular}{cccc}
 \hline
  \hline
 a & q$_{a}$ (erg) & C$_{a}$ & $\eta_{a}$\\
 \hline
 1 &  $6.58566\times10^{-24}$ &  0.005897 & 0.7207\\
 2 & $1.03841\times10^{-23}$ &  0.04886 & 1.9441\\
 3 & $2.39881\times10^{-23}$ &  0.008197 &  5.5444\\
 4 & $3.42764\times10^{-23}$ &  0.03399 & 1.0772\\
 5 &  $5.42343\times10^{-23}$ &  0.04715 & 4.5935\\
 6 & $9.71087\times10^{-23}$ &  0.02723 & 0.4655\\
 7 & $1.1913\times10^{-22}$ &  0.01257 &  3.2566\\
 8 & $1.8903\times10^{-22}$ &  0.004839 & 0.08640\\
 9 &  $3.75299\times10^{-22}$ &  0.009708 & 3.6040\\
 10 & $4.44391\times10^{-22}$ &  0.01036 & 3.6744\\
 11 & $7.44282\times10^{-22}$ &  0.005351 &  2.7812\\
 12 & $1.55073\times10^{-21}$ &  0.01207 & 4.5299\\
 13 &  $2.18166\times10^{-21}$ &  0.006857 & 2.7185\\
 14 & $2.95299\times10^{-21}$ &  0.003468 & 5.9316\\
 15 & $6.24437\times10^{-21}$ &  0.006489 &  3.2129\\
 16 & $8.04399\times10^{-21}$ &  0.002051 & 0.6494\\
 17 &  $1.54649\times10^{-20}$ &  0.003527 & 2.0944\\
 18 & $1.83393\times10^{-20}$ &  0.003387 & 1.5179\\
 19 & $3.44584\times10^{-20}$ &  0.001494 &  1.9025\\
 20 & $6.27323\times10^{-20}$ &  0.0006659 & 4.9010\\
 \hline
 \hline
\end{tabular}
\end{center}
\label{table:20modes}
\end{table}%

As with the case with $N_{k}=4$, we start by identifying the parametric resonances.  By comparing the top panels of figures (\ref{fig:4modes}) and (\ref{fig:20modes}), we see that parametric resonances occur much more frequently with more modes.  Most of these will be eliminated by the density scale height condition.  After performing both steps, we are left with thirty-four potential transition locations.  For clarity in the figure, only those transitions with a predicted transition amplitude greater than 10\% are marked.  Transitions smaller than this are either too small to see on this scale, or they are stopped before they cause any noticeable change.  Six of the transitions are labeled $A$ through $F$ in figure (\ref{fig:20modes}).  For $B$ through $F$, we see evidence of a transition on the order of 30\% or larger.  Figure (\ref{fig:20modesClose}) shows a close up of the last $3\times10^{4}$ km of propagation, with all panels as described before.  In this close up, we see the transitions $E$ and $F$, observed in figure (\ref{fig:20modes}), as well as the remaining transitions now labeled $G$ through $P$.

We note here that some of these transitions are very close to one another, for example $J$ and $K$.  In this case, the transitions actually overlap one another.  Transition $J$ occurs at $r=3.057\times10^{9}$ cm with a predicted wavelength of $1/Q_{J} \sim 5\times10^{7}$ cm.  Transition $K$, meanwhile, occurs at $r=3.101\times10^{9}$ cm with a predicted wavelength $1/Q_{K} \sim 1.8\times10^{8}$ cm.  The distance between the two locations is less than either of the predicted wavelengths.  What we observe is transition $J$ beginning, then being interrupted by transition $K$.   

From these examples, we can observe the effect of adding more modes to the turbulence.  As mentioned, there are many more parametric resonances in the $N_{k}=20$ case than in the $N_{k}=4$ example.  The larger number of modes results in a higher probability that one of the frequencies will correspond to $\delta \breve{k}$.  Also, there are more frequencies to add to one another in combinations to equal $\delta \breve{k}$.  For example, the four parametric resonances located in the $N_{k}=4$ case are all caused by a single mode.  In fact, they are all caused by the $a=1$ mode.  In contrast, in the $N_{k}=20$ case, several of the parametric resonances are from combinations of two modes, and those with only a single mode are not all from the same mode.  The higher number of parametric resonances leads to more transitions overall.  We can expect that with even more modes, this trend will continue.  

With the larger number of transitions, we also see that the locations fall closer together.  This results in the overlapping noted in figure (\ref{fig:20modesClose}) for transitions $J$ and $K$.  The overlap seen in that example can be expected to occur more frequently as the number of modes increases.  Predictions could prove more complicated with very large numbers of modes for this reason.

These two examples show the effectiveness of this two step method for individual neutrinos propagating through specific instances of turbulence. We now apply this method to investigate the effects of changing the parameters of the turbulent spectrum.

\section{Changing Turbulence Parameters}\label{Vary}

\subsection{Effect of RMS Amplitude}\label{CrmsVary}

As mentioned in Section \ref{Methods}, there are two important parameters when defining the turbulent field: $C_{rms}$ and $\lambda_{cut}$.  The first parameter we investigate is the RMS amplitude of the turbulence.  The RMS amplitude, $C_{rms}$, is responsible for setting the overall size of the fluctuations.  A larger $C_{rms}$ will increase the amplitude of the perturbation at every wavelength.  The values reported in previous work have range from well below 1\% to 50\% or more.  Since there is such a wide range, we need to know how these different levels of turbulence affect the parametric resonances.  

We will show that there are two effects of changing $C_{rms}$: a broadening of the resonances, and an increase in the number of parametric resonances.  These two effects will be illustrated using the example in figure (\ref{fig:4modes}).  Recall that the turbulence in figure (\ref{fig:4modes}) used $N_{k}=4$, $C_{rms}=0.1$, and $\lambda_{cut} =314.1$ km.  We will now hold the wavelengths of the four modes constant while varying the RMS amplitude, allowing it to take on the values $C_{rms} = 0.05, \,0.1, \,0.3$, and 0.5.  

\begin{figure}
\begin{centering}
\includegraphics[width=\linewidth]{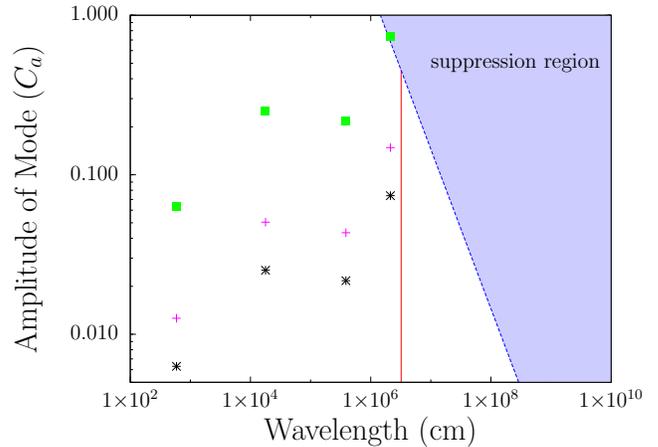}
\caption{Important wavelength scales at the location of transition $B$ in figure (\ref{fig:CrmsVary}).  The values of $\lambda_{fluct,split}$ and $\lambda_{fluct,ampl}$, as well as the suppression region, are indicated as in figure (\ref{fig:lambdaVsC}). The wavelengths and amplitudes of the four modes in the turbulent spectrum are plotted for three values of $C_{rms}$.  $C_{rms} = 0.05$ is shown as black stars, $C_{rms} =0.1$ as magenta crosses, and $C_{rms} = 0.5$ as green boxes.}\label{fig:lambdaVsC4}
\end{centering}
\end{figure}

Before examining the RWA solution and the numeric results, we can first make a prediction using the known important wavelengths discussed in Section \ref{Methods}.  We will focus on a particular location, that of transition $B$ previously identified in figure (\ref{fig:4modes}) which is at a density of $\rho = 874$ g/cm$^{3}$.  Using equations (\ref{eq:lambdaSplit}) and (\ref{eq:FluctAmpl}), as well as the known wavelengths and amplitudes for the turbulence spectrum used in this case, we create a plot similar to that of figure (\ref{fig:lambdaVsC}).  The results of this are shown in figure (\ref{fig:lambdaVsC4}).  The values of $\lambda_{fluct,split}$ and $\lambda_{fluct,ampl}$ are indicated as before by solid red and dashed blue curves, respectively.  The suppression region is shaded and labeled.  We show the wavelengths and amplitudes for the turbulent spectrum for three of the values of RMS amplitude: $C_{rms}=0.05$ as black stars, $C_{rms} = 0.1$ as magenta crosses, and $C_{rms} =0.5$ as green boxes.  

We clearly see that one of the points in this particular spectrum falls almost exactly on $\lambda_{fluct,split}$, indicating a strong possibility of a stimulated transition at this point.  Since the amplitude of this mode is increasing, the transition should increase in amplitude as well.  We can also see that for the highest value of $C_{rms}$ (green boxes) this point actually falls into the suppression region.  In this case, however, we do not expect suppression.  The suppression region and $\lambda_{fluct,ampl}$ are defined for those modes not contributing the parametric resonance, i.e. those with $n_{\star a} =0$.  The mode in this case is nearly equal to $\lambda_{fluct,split}$, and thus contributing to the resonance ($n_{\star a}=1$).  Thus we will still expect a large transition at this location.  

\begin{figure*}[t!]
\begin{centering}
\includegraphics[width=\textwidth]{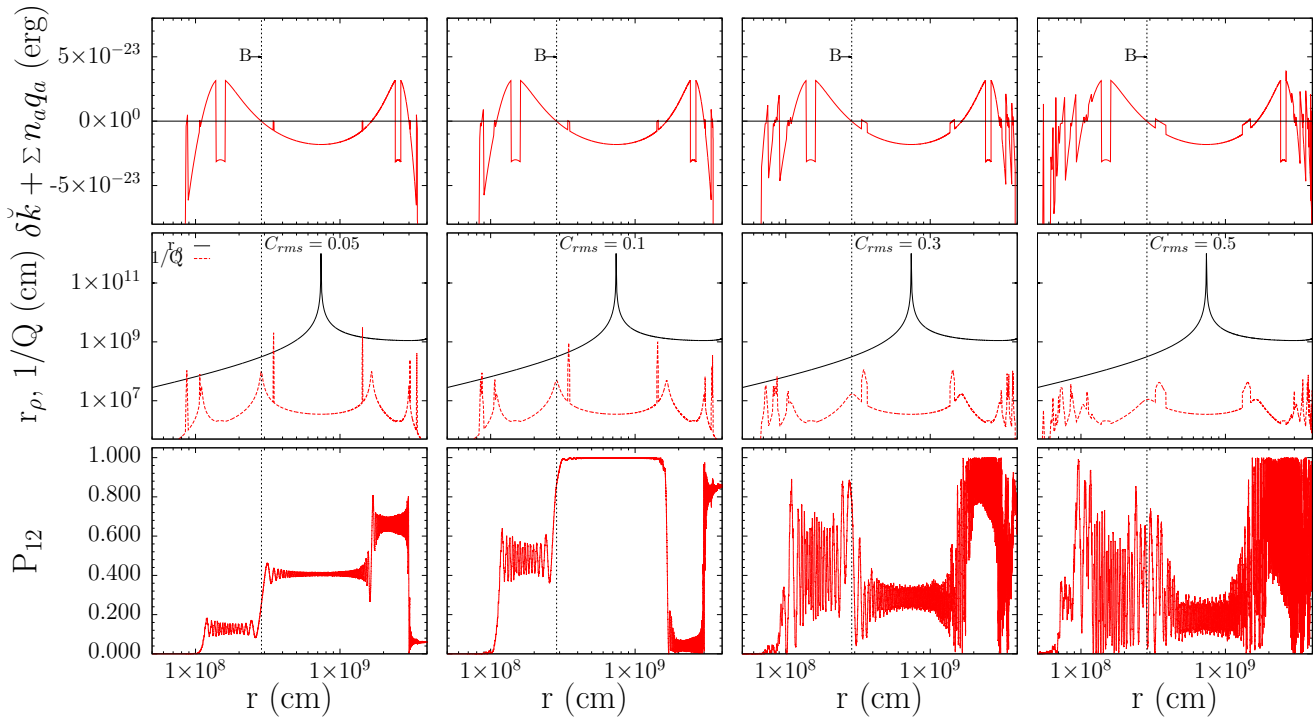}
\caption[Effects of changing the RMS amplitude of the turbulence.]{Effects of changing the RMS amplitude of the turbulence.  Top, middle, and bottom panels are as described for figure (\ref{fig:4modes}).  The RMS amplitude increases from left ($C_{rms}=0.05$) to right ($C_{rms}=0.5$), while the wavelengths of the four modes are held constant.  Transition $B$, previously identified in figure (\ref{fig:4modes}) is indicated by a dashed vertical line for each case.}\label{fig:CrmsVary}
\end{centering}
\end{figure*}

The results of the RWA solution and the numeric calculation can be seen in figure (\ref{fig:CrmsVary}).  As in the previous figures, the top panel shows the locations of parametric resonances, the middle panel displays the predicted wavelengths as well as the density scale height, and the bottom panels show the probability $P_{12}$.  The RMS amplitude increases from left to right.

In each of the four cases shown in figure (\ref{fig:CrmsVary}), similar transitions can be seen.  The transition location $B$, identified in figure (\ref{fig:4modes}), is labeled and indicated with dashed vertical lines.  With the relatively low values of $C_{rms}=0.05$ and 0.1, this transition is very clearly distinguished.  In the latter two panels, with $C_{rms}=0.3$ and 0.5, this location still shows a transition, but it is not as sharp.  We can also see that the height of all of the transitions increases as the RMS amplitude increases so that, for example, the transition $B$ is approximately 30\% with $C_{rms}=0.05$ and nearly 80\% with $C_{rms}=0.5$.  The wavelengths predicted in the small $C_{rms}$ case are longer than the transition regions, resulting in the transitions being cutoff before reaching 100\%.

This progression can be explained using the RWA solution.  If we examine the middle panel for the $C_{rms} = 0.05$ case, we see that the peak for transition $B$ indicated by the dashed vertical line is fairly narrow.  The system spends a relatively short time at the peak wavelength and high amplitude, causing a transition of less than 100\%. As the value of $C_{rms}$ increases, we see this peak broaden while the maximum value simultaneously decreases.  In this situation the wavelength of the transition is less than the distance over which the transition occurs, allowing a nearly complete oscillation from one state to the other.  The broadening and shortening of the resonance peaks is the first effect of increasing the RMS amplitude.  
 
Looking at the top panels in figure (\ref{fig:CrmsVary}), we also see that as $C_{rms}$ increases there is an increase in the number of points where the parametric resonance condition is met, as evidenced by the increase in intersections between the black and red curves.  At first glance, this is puzzling, since the wavelengths of the various modes have not changed.  The difference comes about because the sets of RWA integers $n_{\star a}$ change as the RMS amplitude increases. For larger RMS amplitudes combinations of modes become more important than having a single mode with $q_a \sim \delta \breve{k}$.  This is the second effect of increasing the value of $C_{rms}$.  

We can understand this through the RWA solution itself.  The RWA solution combines the Fourier modes of the turbulence in such a way that $|\delta\breve{k} + \sum_{a}n_{a}q_{a}|$ is minimized while the transition amplitude $\kappa^{2}/Q^{2}$ is maximized.  The combinations that achieve both of these goals will change as the amplitudes of the modes change.  For instance, in some cases a combination that is further from the parametric resonance is actually better because it produces a higher amplitude transition.   

Increasing the amplitude of the underlying density perturbations decreases the wavelength $1/Q$ and increases the amplitude $(\kappa/Q)^2$ of the parametric resonance transitions.  Mathematically this comes about through the $J_{n \neq 0}(z_{a})$ terms becoming larger as $z_{a}$ increases.  Thus as you increase $C_{a}$, more opportunities open up for parametric resonance transitions. This is seen in the top panel of Fig. (8).  

\begin{table*}[t!]
\caption{Comparison of the $n_{\star a}$ values chosen at a particular point ($r = 6.7\times10^{7}$ cm) for different values of $C_{rms}$, as shown in figure \ref{fig:CrmsVary}.  Also included are the values for $z_{a}$, the values of the Bessel functions $J_{n_{\star a}}(z_{a})$, the sum $|\delta\breve{k} + \sum_{a}n_{a}q_{a}|$ and the RWA amplitude $\kappa^{2}/Q^{2}$.}
\begin{center}
\begin{tabular}{ccccc}
\hline
\hline
 & $C_{rms} = 0.05$ & $C_{rms} = 0.1$ & $C_{rms} = 0.3$ & $C_{rms} = 0.5$ \\
 \hline
n$_{\star 1}$ & 0 & 0 & 0 & 1\\
n$_{\star 2}$ & 1 & 1 & 2 & 4\\
n$_{\star 3}$ & 0 & 0 & 0 & 0\\
n$_{\star 4}$ & 0 & 0 & 0 & 0\\
\hline
z$_{1}$ &$-0.5373$ & $-1.0747$&$-3.2240$ & $-5.3734$\\
z$_{2}$ &$-0.24771$ & $-0.4895$&$-1.4686$ & $-2.44771$\\
z$_{3}$ &$-3.800\times10^{-4}$ & $-7.601\times10^{-4}$&$-2.280\times10^{-3}$ & $-0.003800$\\
z$_{4}$ &$-1.665\times10^{-4}$ & $-3.330\times10^{-4}$&$-9.991\times10^{-3}$ & $-0.001665$\\
\hline
$J_{n_{\star 1}}(z_{1})$ &$0.9291$ & $0.7314$&$-0.3264$ & $-0.0504$\\
$J_{n_{\star 2}}(z_{2})$ &$-0.12147$ & $-0.2375$&$-0.5534$ & $-0.5096$\\
$J_{n_{\star 3}}(z_{3})$ &$1$ & $1$&$0.9999$ & $0.9999$\\
$J_{n_{\star 4}}(z_{4})$ &$1$ & $1$&$1$ & $0.9999$\\
\hline
$|\delta \breve{k} + \sum_{a}n_{\star a}q_{a}|$ (erg) &1.00062$\times10^{-22}$ & 1.00062$\times10^{-22}$& 6.484$\times10^{-23}$ & 9.1863$\times10^{-24}$ \\
$\kappa^{2}/Q^{2}$ &3.030$\times 10^{-5}$ &7.178$\times10^{-5}$ & 1.214$\times10^{-4}$ & 2.675$\times10^{-3}$ \\
\hline
\hline
\end{tabular}
\end{center}
\label{table:CstarVary}
\end{table*}%

In the case of $C_{rms} = 0.05$, the small amplitudes $C_{a}$ for each mode result in small values for $z_{a} \sim C_{a}V_{0}/\hbar c q_{a}$, the argument of the Bessel functions used in equation (\ref{eq:kappa}) to calculate $\kappa$.  The value of the Bessel function $J_{0}$ is much greater than that of any other $J_{n}$ at small $z_{a}$.  Since $\kappa$ is a product of these Bessel functions, the best solution is for the single mode closest to $q_{a}\sim \delta \breve{k}$ to have $n_{\star a}=1$ while the other modes have $n_{\star a}=0$.  In fact, this is the type of solution chosen for almost the entire propagation distance in the $C_{rms}=0.05$ example.  

As $C_{rms}$ increases, the amplitudes $C_{a}$ and thus the values for $z_{a}$ will also increase.  As the $z_{a}$ values move away from zero, Bessel functions other than $J_{0}(z_{a})$ will become important.  The best solution for the RWA is more likely to be a combination of two modes with $n_{\star a}\neq 0$.  This proves true for the RWA predictions for the cases with $C_{rms}=0.3$ and 0.5.  The extra parametric resonance locations seen in the top panels of figure (\ref{fig:CrmsVary}) are all met through combinations of multiple modes.  This mathematical explanation shows that as $C_{rms}$ increases, the importance of having single modes with $q_{a} \sim \delta\breve{k}$ decreases.  

Table (\ref{table:CstarVary}) shows the changing values of $n_{\star a}$ at a the point $r=6.7\times10^{7}$ cm.  This location corresponds to a parametric resonance only when $C_{rms} =0.5$.  We can see that as $C_{rms}$ increases, the solution changes from a single mode with $n_{\star a} \neq 1$ to a combination of modes.  While the value of $|\delta \breve{k} + \sum_{a}n_{\star a}q_{a}|$ is larger for the small values of $C_{rms}$, changing the values of $n_{\star a}$ will result in lower predicted amplitudes due to the values of $z_{a}$ and the Bessel functions in the definition of $\kappa$.

The increase in the number of resonances results in overlapping of transitions, as we saw in figure (\ref{fig:20modesClose}).  The overlapping of transitions leaves us with a more chaotic looking probability.  If we continued to increase $C_{rms}$, we would eventually lose any recognizable features.

\subsection{Effect of Length Scale Cutoff}\label{lambdaVary}

The other main parameter defining the turbulence spectrum is the cutoff scale $\lambda_{cut}$.  This parameter controls the longest wavelengths allowed in the turbulent spectrum.  In addition, it controls the amount of power for the shorter wavelength modes in the spectrum.  For example, if $\lambda_{cut}$ is increased from 300 km to 3000 km while holding $C_{rms}$ constant, the amplitudes for modes with wavelengths $\sim$10 km will decrease.  Previous work has used cutoff wavelengths that vary by orders of magnitude from one study to the next \cite{Reid,Burgess, Sawyer, 2013PhRvD..88b3008L, Kneller:2010sc, 2013arXiv1302.3825K, 2013PhRvD..88d5020K}.  This wide range and our knowledge of the wavelengths which most affect neutrino transformations necessitates an investigation into how these different cutoff scales change the neutrino propagation in the SN model.  

We described a suppression effect due to the long wavelength modes present in a turbulent spectrum in Section \ref{Methods}.  To study the effects of varying $\lambda_{cut}$, we start with the $N_{k}=20$ example shown in figure (\ref{fig:20modes}).  Keeping the value of $C_{rms}$ constant at 0.05, we increase the value of $\lambda_{cut}$ by factors of 10.  In effect, this increases the wavelength of each mode by a factor of 10.  Once again, this has a dual effect on the spectrum.  It not only allows longer wavelengths, which can cause suppression, but decreases the amplitudes of the shorter wavelength modes as well.  

\begin{figure}[ht!]
\begin{centering}
\includegraphics[width=\linewidth]{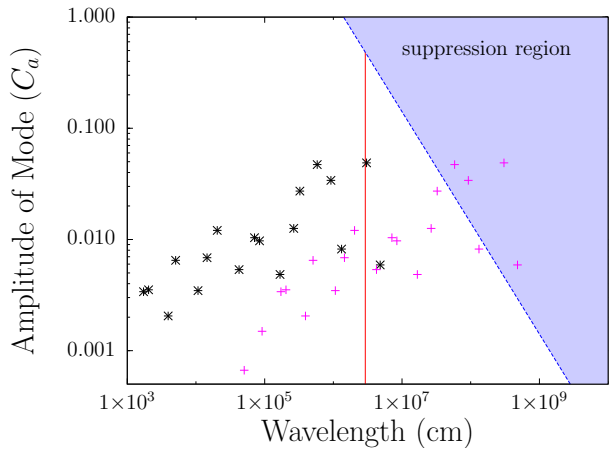}
\caption{Important wavelength scales at the location of transition $E$ in figure (\ref{fig:kCutVary}).  The values of $\lambda_{fluct,split}$ and $\lambda_{fluct,ampl}$, as well as the suppression region, are indicated as in figure (\ref{fig:lambdaVsC}). The wavelengths and amplitudes of the twenty modes are shown for two values of $\lambda_{cut}$.  Black stars are used for $\lambda_{cut} = 314.1$ km and magenta crosses for $\lambda_{cut} = 314100$ km.}\label{fig:lambdaVsC20}
\end{centering}
\end{figure}

Once again before studying the results of the RWA solution and numerical calculation, we can create a figure of the important wavelengths and amplitudes and how this particular turbulent spectrum compares.  This is shown in figure (\ref{fig:lambdaVsC20}).  We have chosen to focus on a particular transition, this time the transitions previously identified as $E$ in figure (\ref{fig:20modes}).  The values of $\lambda_{fluct,split}$ and $\lambda_{fluct,ampl}$, as well as the suppression region, are indicated as in earlier figures.  The points now show the wavelengths and amplitude for the original turbulent spectrum with $\lambda_{cut} = 341.1$ km in black stars, and the values with $\lambda_{cut}$ increased by a factor of 100 in magenta crosses.  

From this figure, we see that some of the points for the original spectrum lie close to the value of $\lambda_{fluct,split}$ and none fall within the suppression region.  We would therefore expect a large transition at this point, exactly as was observed previously.  In contrast, the modified spectrum with a longer cutoff wavelength has several points inside the suppression region.  Based on this, we should expect the transition at this location to be suppressed for the larger value of $\lambda_{cut}$.  

\begin{figure*}[ht!]
\begin{centering}
\includegraphics[width=\linewidth]{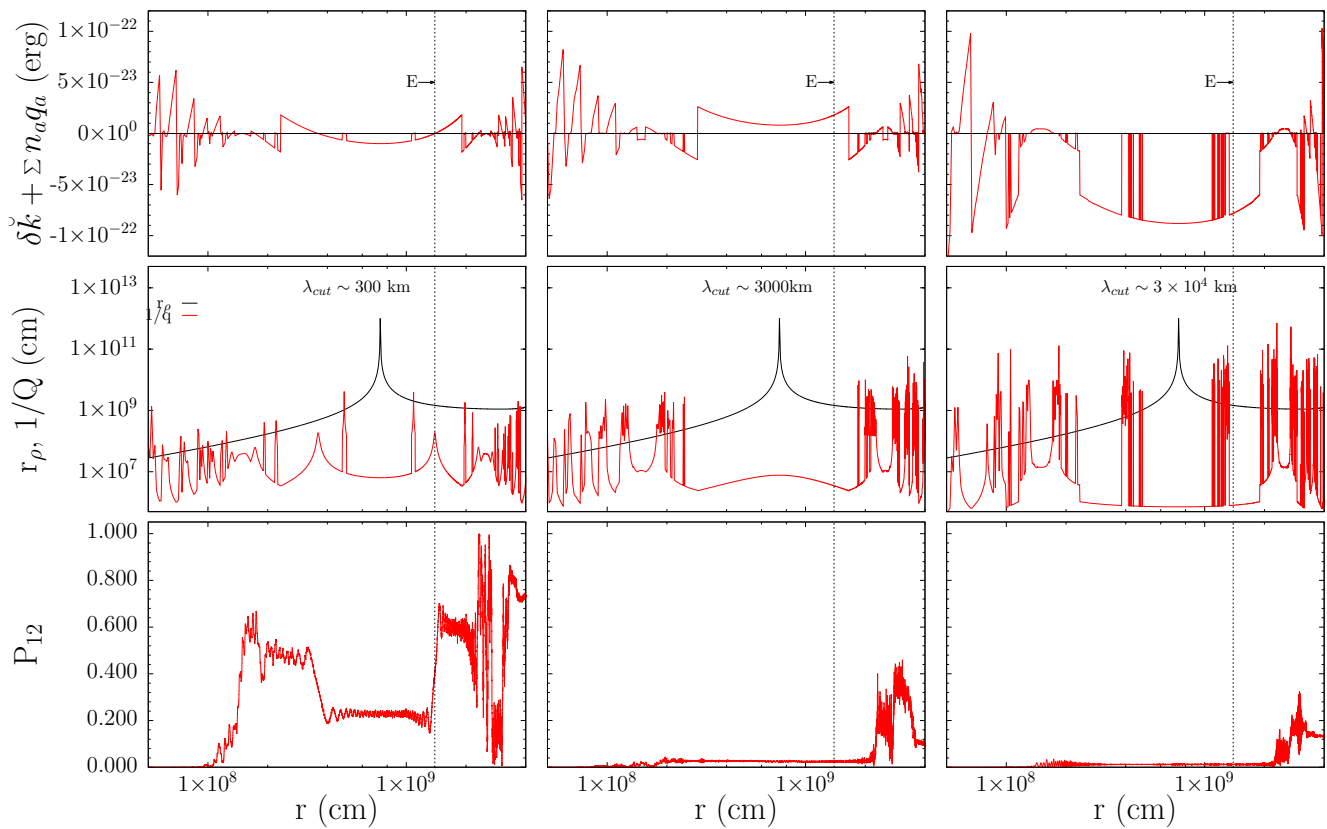}
\caption[Effects of varying the cutoff scale of the turbulence.]{Effects of varying the cutoff scale $\lambda_{cut}$ of the turbulence. Top, middle, and bottom panels are as described for figure (\ref{fig:4modes}).  The cutoff scale is increased by a factor of 10 for each column.  A clear suppression in the probability $P_{12}$ is visible as the cutoff scale increases. Transition $E$, located in figure (\ref{fig:20modes}) is identified by a dashed vertical line in each case. }\label{fig:kCutVary}
\end{centering}
\end{figure*}

The results of this study can be seen in figure (\ref{fig:kCutVary}).  The top, middle, and bottom panels are the same as described above.  The cutoff scale is increased from left to right, from $314.1$ km in the leftmost panels to $31410$ km in the rightmost.  By simply examining the bottom panels, we see a marked suppression as the cutoff scale is increased.  The location chosen for study in figure \ref{fig:lambdaVsC20} is indicated by the dashed vertical line labeled $E$.  

We can explain this suppression through both long wavelength suppression and a lengthening of predicted wavelengths.  Adding long wavelengths to the turbulent spectrum will cause suppression in approximately half of cases \cite{Patton:2013}.  The suppression is strongest when the ratio of the amplitude and frequency of the mode, given by the parameter $z_{a} \sim C_{a}V_{0}/\hbar c q_{a}$, are on the order of the zeroes of the Bessel function $J_{0}$, the first of which occurs at $z_{a}\approx 2.4$.  By allowing longer wavelengths in the spectrum, we increase the opportunity for attaining this ratio.  In fact, the highest value of $z_{a}$ attained in the case with $\lambda_{cut} = 314.1$ km is approximately 0.5, while in the case of $\lambda_{cut} = 31410$ km the first five modes have $z_{a}>1$ for most of the propagation.  These first five modes would fall within or very close to the shaded blue suppression region figure (\ref{fig:lambdaVsRho}).  

We also know that increasing the cutoff scale reduces the power for the shorter wavelengths.  Among these wavelengths are those that correspond to the mass splitting scale, which are the ones that cause the parametric resonance.  These modes have $n_{\star a}=1$ in the RWA solution. From equation (\ref{eq:kappa}), we know that $\kappa \propto \prod_{a} J_{n_{\star a}}(z_{a})$.  As the amplitude of the $n_{\star a} = 1$ modes decrease, the values of $J_{1}(z_{a})$ and thus of $\kappa$ will also decrease.  Since at parametric resonance $\kappa^{2} = Q^{2}$, this results in a longer predicted wavelength, well above the density scale height.  As such, reducing the power in the shorter wavelength modes causes a suppression in addition to that caused by the long wavelength modes.  

From these results, we can conclude that turbulence which follows a Kolmogorov spectrum and has a long cutoff scale will display some suppression of transitions.  This information can be used to predict the possible neutrino transitions in a density produced from a SN simulation, before performing a numerical calculation.  

\section{Stimulated Transition in Antineutrinos}\label{AntiNu}

All of our previous results have been for neutrinos in the normal hierarchy.  For this hierarchy in the standard MSW picture, transitions between antineutrino species are not expected at any point along the profile. Parametric resonance, which relies on matching frequencies instead of density scales, allows for possible antineutrino transitions.  Here, we use the methods discussed above to search for these transitions.

To make the switch from neutrinos to antineutrinos, the sign of the matter potential will change from positive to negative, such that $H_{\bar{\nu}}^{(f)} = U_{0}K^{(m)}U_{0}^{\dag} - V^{(f)}$.  All other aspects of the calculation will remain the same.  Due to this change in the Hamiltonian, the eigenvalue splitting for the antineutrinos $\breve{\overline{\delta k}}$ will not be the same as for the eigenvalue splitting for the neutrinos. As such, the locations of parametric resonances for antineutrinos are not expected to be the same as those for neutrinos.  

\begin{figure}
\includegraphics[width=\linewidth]{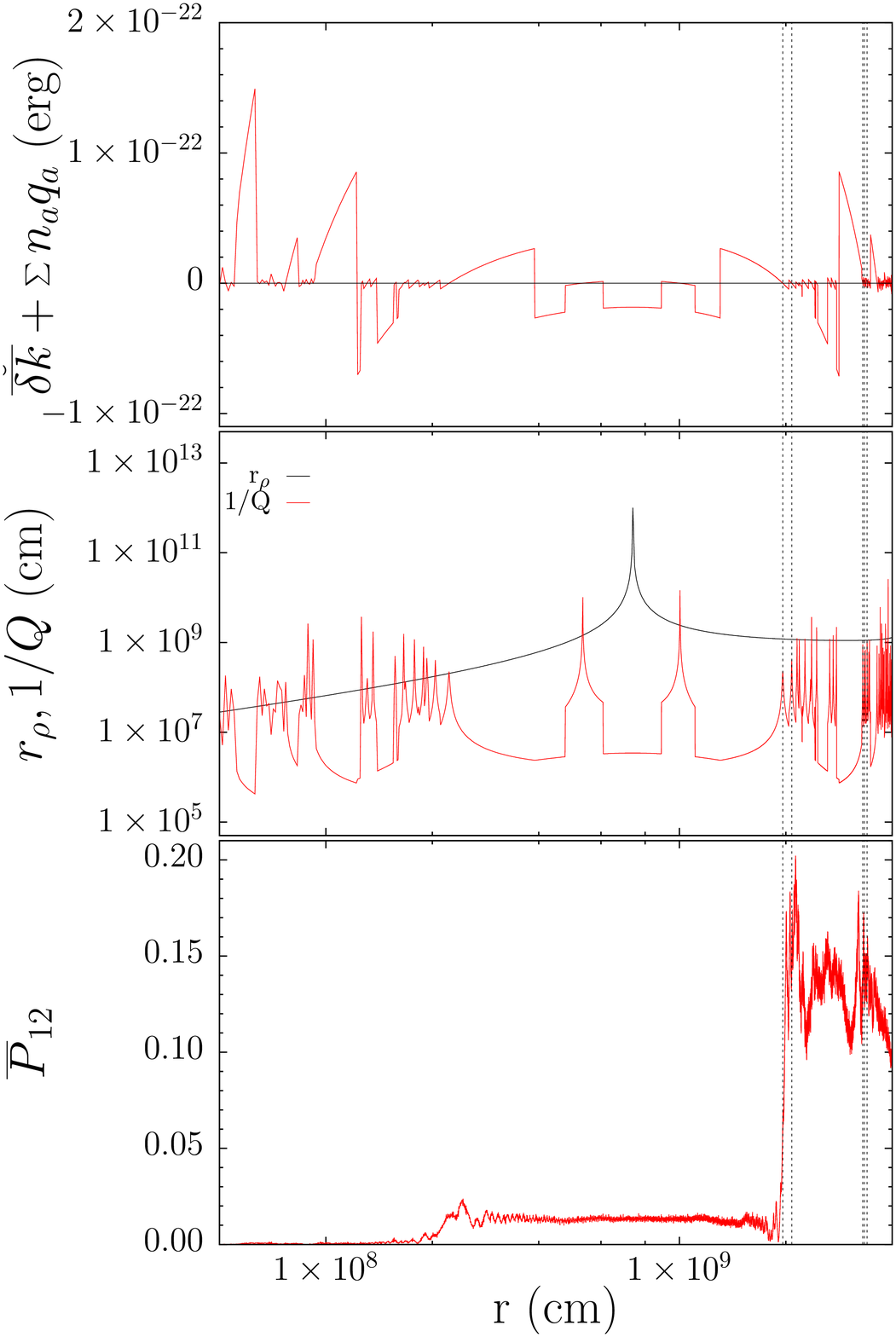}
\caption{Same as figure (\ref{fig:4modes}), but for antineutrinos.  The perturbation in this case has $N_{k}=20$, $C_{rms} = 0.1$, and $\lambda_{cut} = 314.1$ km.}\label{fig:AntiNu}
\end{figure}

As with the neutrinos in previous sections, we begin by searching for locations where the parametric resonance condition is met.  Next, we check to see if the predicted wavelengths at those locations fall below the density scale height.  Finally, we look at a numerically generated $\overline{P}_{12}$ and compare locations of actual transitions to the predictions.  This can be seen in figure (\ref{fig:AntiNu}), which was created using a Kolmogorov spectrum with 20 modes, $C_{rms} = 0.1$, and $\lambda_{cut} = 314.1$ km.  While this is a similar set of turbulence parameters as used for figure (\ref{fig:20modes}), this is a completely different set of wavelengths and amplitudes. The values used are listed in table (\ref{table:antiNu}). 

\begin{table}[t]
\caption{Values of $q_{a}$, $C_{a}$ and $\eta_{a}$ used for the $N_{k}$ = 20 modes example shown in figure \ref{fig:AntiNu} with $\lambda_{cut} = 314.1$ km and $C_{rms} = 0.1$.}
\begin{center}
\begin{tabular}{cccc}
 \hline
  \hline
 a & q$_{a}$ (erg) & C$_{a}$ & $\eta_{a}$\\
 \hline
 1 &  $7.60623\times10^{-24}$ &  0.02010 & 2.4651\\
 2 & $1.50898\times10^{-23}$ &  0.03096 & 4.1117\\
 3 & $1.84269\times10^{-23}$ &  0.02776 &  5.9080\\
 4 & $3.63138\times10^{-23}$ &  0.009348 & 4.5564\\
 5 &  $6.04137\times10^{-23}$ &  0.006186 & 0.6541\\
 6 & $7.2492\times10^{-23}$ &  0.01866 & 3.6416\\
 7 & $1.19595\times10^{-22}$ &  0.02434 &  4.5985\\
 8 & $2.4881\times10^{-22}$ &  0.004374 & 2.3538\\
 9 &  $3.82756\times10^{-22}$ &  0.005666 & 1.8986\\
 10 & $4.32779\times10^{-22}$ &  0.01375 & 5.3892\\
 11 & $7.01636\times10^{-22}$ &  0.003325 & 0.7955\\
 12 & $1.136\times10^{-21}$ &  0.002712 & 0.3970\\
 13 &  $1.64156\times10^{-21}$ &  0.0002635 & 3.3373\\
 14 & $2.77841\times10^{-21}$ &  0.0001231 & 5.5685\\
 15 & $6.31359\times10^{-21}$ &  0.008661 &  2.8629\\
 16 & $7.51972\times10^{-21}$ &  0.001621 & 4.6584\\
 17 &  $1.04759\times10^{-20}$ &  0.005016 & 2.4274\\
 18 & $1.70742\times10^{-20}$ &  0.003825 & 2.4564\\
 19 & $3.33396\times10^{-20}$ &  0.002563 &  5.1533\\
 20 & $5.8992\times10^{-20}$ &  0.001001 & 3.4448\\
 \hline
 \hline
\end{tabular}
\end{center}
\label{table:antiNu}
\end{table}%

Figure (\ref{fig:AntiNu}) has the same three types of panels as figures (\ref{fig:4modes}) - (\ref{fig:kCutVary}).  Once again for clarity in the figure, only transitions with a predicted amplitude larger than 10\% are indicated by vertical dashed lines. 

In the top panel, we can observe many points where the parametric resonance condition is met.  However, nearly all of them are eliminated by the density scale height condition as can be seen in the middle panel.  The bottom panel shows $\overline{P}_{12}$, where we see that the predicted locations for transitions line up well with actual features in the probability.  In particular, a transition of $\sim 20\%$ is seen near 2$\times10^{9}$ cm. 

The size of transitions for antineutrinos are typically of smaller amplitude even when $C_{rms}$ is the same, as can be seen by a direct comparison of the bottom panels of figure (\ref{fig:AntiNu}) and figures (\ref{fig:4modes}) and (\ref{fig:20modes}).  Similar to the case of small $C_{rms}$ discussed above and shown in the leftmost panel of figure (\ref{fig:CrmsVary}), the resonances for the antineutrinos are very narrow.  This causes the system to spend relatively little time on the resonance, resulting in incomplete transitions.

The narrow resonances can be explained by the fact that the antineutrino system is far from the MSW resonance.  As mentioned in \cite{Kneller:2012id}, the width of the resonance is proportional to $|\kappa|$.  From equation (\ref{eq:kappa}), we see that $|\kappa| \propto |\breve{U}_{e1}^{\star} \breve{U}_{e2}|$.  This product of mixing matrix elements is small when far from the MSW resonance, resulting in a small value of $|\kappa|$ and a correspondingly narrow resonance.    

This example clearly illustrates the difference between the MSW and parametric resonance viewpoints.  Using strictly MSW methods, no transitions of any size would be predicted for antineutrinos in the normal hierarchy.  Using the parametric resonance method with the RWA solution, we are able to predict that in fact some transitions should occur as well as locate them as we did for the neutrinos.  

\section{Application to Hydrodynamical SN Simulations}\label{Application}

A recent paper by Borriello \textit{et al.} \cite{borriello} performed a Fourier analysis on the results of a 2D SN simulation from the Garching group \cite{Kifondis:2003, Kifondis:2006}. In that work, analysis was carried out for 768 radial directions.  For each of these directions, average densities and complete power spectra were found for the section of the profile before the forward shock.  Neutrino crossing probabilities were then calculated for the entire profile, including the shocks.  The authors concluded that the turbulence had no real effect on the neutrinos, and the most relevant piece of the density profile is in fact the shock.

\begin{figure*}
\includegraphics[width=\linewidth]{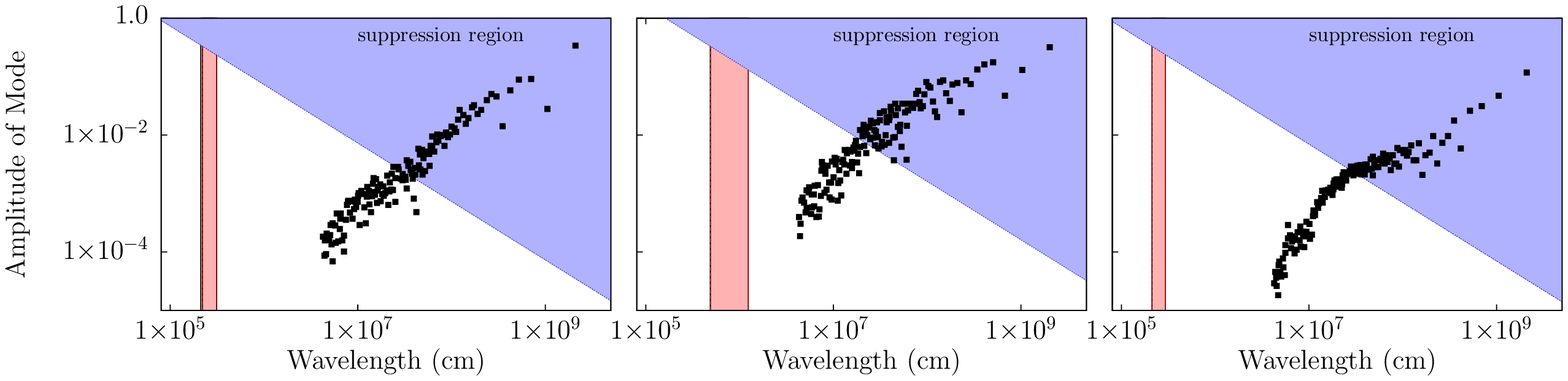}
\caption{Amplitudes and wavelengths (black points) from a 2D SN simulation, as reported in \cite{borriello}.  The suppression region is shaded blue, with $\lambda_{fluct,ampl}$ defined by a dashed blue line.  We have now plotted a parametric resonance region in red for a range of energies from 5 MeV to 50 MeV.  }\label{fig:Borriello}
\end{figure*}

We have examined the results in order to determine whether it is possible to explain the results using the RWA approach.  Three specific cases from the 768 total radial directions were used by Borriello \textit{et al.} as examples, with the density profile and power spectra shown.  Using these three examples, we have determined the approximate average density, wavelengths and amplitudes for the the three radial directions used in that work.  From this information, we can create a plot of wavelength and amplitude like figure (\ref{fig:lambdaVsC}).  This is shown in figure (\ref{fig:Borriello}).

Unlike in figure (\ref{fig:lambdaVsC}) which showed the relevant wavelengths for a single neutrino energy, here a red band is used to represent $\lambda_{fluct,split}$ for a range of energies from 5 to 50 MeV.  The values of $\lambda_{fluct,ampl}$ are also plotted for various energies in blue dashed lines, but the different energies are indistinguishable on the scale of these plots.  Black points represent the wavelengths and amplitudes found in the 2D SN analysis.

The first thing to notice is that none of the points for wavelength and amplitude of these spectra fall within the parametric resonance region indicated by the red band.  As such, we can already expect that these profiles will not produce large transitions due to parametric resonance for any energy of neutrino.  In addition, many of these amplitudes and wavelengths fall within the suppression regions.  These modes are likely further suppressing any small transitions that might have occurred.  Based on this analysis, we would also expect to see very little occurring in the region where turbulence is present, exactly as observed by Borriello \textit{et al.} \cite{borriello}. 

In order to produce stimulated transitions, these spectra could be modified in several ways.  The first would be to reduce the length of the shortest wavelengths, pulling the lower edge of the spectrum into the parametric resonance region.  The length is constrained by the resolution of this particular SN simulation, which in this case was on the order of 10 km, but the resolution of these type of simulations is constantly improving.  

In addition, strong stimulated transitions are being suppressed by the many modes falling in the suppression region.  As such, even if the spectra are modified to include shorter wavelengths, the size of any transitions will be diminished.  These modes must be removed to eliminate the suppression effect.  This would obviously be an artificial process, as those long wavelength modes were observed in simulated SN data.  

Our analysis using the RWA method and the knowledge of the important scales of the problem show that no stimulated transitions should be expected in the turbulence region of the SN simulation.  This matches well with the conclusions of the authors.  Applying our method to simulation data and obtaining similar results shows the strength of the analytic solution we have presented here.

\section{Conclusions}

We have extended the work of \cite{Kneller:2012id} and \cite{Patton:2013} to encompass a changing density profile.  Using the concept of parametric resonances, we have developed a method to predict the locations of possible transitions in a model SN profile.

From the RWA solution, we locate points where the parametric resonance condition is fulfilled.  If the predicted wavelength at these points is lower than the density scale height, a transition is expected.  We have shown that this simple method can pinpoint the transitions that appear in the probability. We have shown that this method accurately predicts the locations of large transitions, both in an example with $N_{k}=4$ modes and one with $N_{k}=20$ modes. 

We have also shown the effects of changing the RMS amplitude and cutoff scale of the turbulence.  Increasing the RMS amplitude causes two effects: a higher number of parametric resonances and a decrease in the transition wavelengths.  The resonance peaks are also broader, meaning that the system spends a longer amount of time at the shorter wavelength, allowing for complete transitions from one state to another.  The increase in resonances, along with the shorter transition wavelengths, causes many overlapping transitions for higher RMS amplitudes.   The end result is increasingly chaotic looking probabilities as the RMS amplitude increases.  

Increasing the cutoff scale of the turbulence also results in two simultaneous effects.  First, the long wavelength modes cause a suppression, as described in \cite{Patton:2013}.  These modes cause a distortion to the system, to the point that a parametric resonance no longer causes a transition.  This suppression reduces the amplitude of any transitions that do still occur.  Second, increasing the cutoff scale lowers the amplitudes of short wavelength modes, some of which cause the parametric resonance.  This second effect causes longer transition wavelengths, so that the wavelength is much longer than the distance over which the transition occurs.  These two effects, the suppression and the long transition wavelengths, cause transitions that are smaller in amplitude and lower in number.

The RWA method has been successfully applied to antineutrinos, demonstrating its usefulness in cases where MSW transitions are not expected.  Using the same criteria described for locating stimulated transitions for neutrinos, we are able to predict where such events would occur in the antineutrinos as well.  These predictions were verified upon inspection of the numerically calculated probabilities.

Finally, we have applied our method to the analysis of a 2D SN simulation \cite{borriello}.  The authors of that study found that the turbulence present had no effect on the flavor transformations of the neutrinos.  We have shown that the wavelengths and amplitudes found in that analysis do not fall within the parametric resonance region, due to the resolution scale being longer than the wavelengths $\lambda_{fluct,split}$.  In addition, many of the modes fall in the suppression region. From these two facts we conclude, as Borriello \textit{et al.} did, that the neutrinos would not undergo any transitions in the turbulence region.  If this is true of all SN, that there are no fluctuations on the scale of $\lambda_{fluct,split}$ and significant fluctuations on the scale of $\lambda_{fluct,ampl}$, then transitions in SN due to parametric resonance are no expected.  Future simulations must be carefully analyzed with an eye toward these important scales.

We have described the important wavelength scales for neutrinos traveling through turbulence, both those that stimulate transitions and those that suppress them.  We have shown that the RWA solution can be used in a problem with a changing density profile, and discussed the effects of changing certain parameters of the turbulence.  We have also been able to apply our method to a separate SN simulation analysis and explain the results found by the original authors.  These steps together show that the RWA solution is an important tool for the future study of neutrino oscillations in turbulence.


\acknowledgments

This work was supported by DOE grants DE-SC0004786 (GCM+JPK+KMP) and DE-SC0006417 (JPK),
DE-FG02-02ER41216 (GCM+KMP) and by 
a NC State University GAANN fellowship (KMP).


\end{document}